\documentstyle[preprint,aps]{revtex}
\newcommand{\bq}{\begin{equation}}
\newcommand{\eq}{\end{equation}}
\begin{document}
\draft
\title{
Role of Layering Oscillations at Liquid Metal Surfaces in
Bulk Recrystallization and Surface Melting}
\author{Orio Tomagnini $^{\rm (1)}$,
        Furio Ercolessi $^{\rm (2,3)}$,
        Simonetta Iarlori $^{\rm (1)}$,
        Francesco D. Di Tolla $^{\rm (2,3)}$, and
        Erio Tosatti $^{\rm (2,3,4)}$}
\address{
$^{\rm (1)}$ IBM--ECSEC, Viale Oceano Pacifico 171/173, I-00144 Rome, Italy}
\address{
$^{\rm (2)}$ Istituto Nazionale di Fisica della Materia (INFM), Italy}
\address{
$^{\rm (3)}$ International School for Advanced Studies (SISSA-ISAS),
Via Beirut 4, I-34014 Trieste, Italy}
\address{
$^{\rm (4)}$ International Center For Theoretical Physics (ICTP),
Trieste, Italy}
\date{1 August 1995}

\maketitle

\begin{abstract}
The contrasting melting behavior of different surface
orientations in metals can be explained in terms of a repulsive or attractive
effective interaction between the solid-liquid and the
liquid-vapor interface.
We show how a crucial part of this interaction
originates from the layering effects near the liquid metal surface.
Its sign depends on
the relative tuning of layering oscillations to the crystal
interplanar spacing, thus explaining the orientational dependence.
Molecular dynamics recrystallization simulations of Au surfaces provide 
direct and quantitative evidence of this phenomenon.
\end{abstract}

\narrowtext

Surface melting (SM), i.e. complete wetting of a crystal surface by a thin
film of its own melt, is very common in nature.
For example, all faces of a real system, like rare gas solid Ar \cite{argon},
as well as of its popular model, the fcc Lennard-Jones (LJ) crystal
\cite{LJ}, are believed to melt completely.
On the other hand, most metals have at least one close-packed face
which does not melt, and is crystalline right up
to $T_m$ \cite{vanderveen}.
This non-melting (NM) behavior is often explained phenomenologically,
in terms of a low solid--vapor (SV)
surface free energy $\gamma_{\rm SV}$, which in this case
can be lower than the sum of solid--liquid (SL)
plus liquid--vapor (LV) interface free
energies $\gamma_{\rm SL}$ + $\gamma_{\rm LV}$.
Open faces, by contrast, tend to have
a higher energy when solid and this favors
SM, in agreement with observations.
At a more microscopic level, Chernov and Mikheev \cite{chernov}
made the interesting additional remark
that surface non-melting could in fact be
ascribed to large minima in a strong oscillatory
{\em effective} interaction
between the SL and the LV interface.
This oscillatory interaction is caused by a layering effect:
the solid (similarly to a hard wall) induces in the liquid film
a layer-like density oscillation with periodicity $2\pi/Q_\circ$
[$Q_\circ$ is the wavevector where the liquid structure
factor $S(Q)$ has its strongest peak], which in turn generates the
interaction. Binding of the two interfaces
in the deepest oscillation at $\ell=0$ is then responsible for a
minimum of $\gamma_{\rm SV}$ and leads to NM.
Alternatively, if the oscillations are weak,
they can be washed out
by capillary waves, the two interfaces unbind and there will be SM.

The purpose of this Letter is to argue theoretically, and to demonstrate
through detailed simulations, that in non-LJ systems, particularly
{\em in metals}, there are in reality not one,
but two distinct layering oscillations---one
with periodicity $2\pi/Q_\circ$, tied to the LV interface,
and one with the periodicity of the interlayer spacing $a$,
tied to the SL interface---which overlap and interfere
inside a liquid metal film.
Tuning and detuning of the two oscillations
depends on orientation, and may dramatically change
the strength of the interaction, thus affecting deeply the surface
melting behavior. In particular, while as expected
well-tuned layering effects disfavor surface melting
(and can make the close-packed surface
very resistant to overheating), {\em the same does not happen for
large detuning}, where SM is favored.

We will proceed in three stages.
First, we give a qualitative, analytical argument showing
that layering effects are strongly dependent on the orientation.
Second, we make this more concrete
through a model 1D wetting calculation  \cite{vicente}
adapted to our case.
Lastly, we present detailed surface recrystallization molecular dynamics (MD)
non-equilibrium simulations which illustrate realistically
the effect of attractive layering forces for a fcc(111) NM face,
in contrast with a fcc(110) SM face where repulsion dominates.

Consider a liquid film of thickness $\ell$, wetting its own solid.
The effective SL--LV interface interaction free energy can be written as
$f(\ell) = f_{\rm SR}(\ell) + f_{\rm L}(\ell)$
where $f_{\rm SR}$ is a short-range effective interaction,
accounting for the merging of the two interfaces
at close contact, and $f_{\rm L}$ is a residual part due to
layering effects, which can be characterized as follows.
The underlying solid propagates into the liquid film
a first density oscillation roughly of the form
$\delta\rho_{\rm SL}(z) \approx 
  k_{\rm S} \exp\left[-(z-z_{\rm SL})/\xi_{\rm S}\right]
\cos\left( 2\pi z/a \right)$
which has the periodicity of the (face-dependent) interlayer
spacing $a$, and decays {\em outwards}
from the SL interface center $z_{\rm SL}$ into the liquid
($z$ grows going from solid to liquid to vapor) within some
characteristic length $\xi_{\rm S}$ \cite{vicente}.
The LV interface of a metal originates however
a {\em second} density oscillation
$\delta\rho_{\rm LV}(z) \approx 
  k_{\rm V} \exp\left[(z-z_{\rm LV})/\xi_{\rm V}\right]
\cos\left[ \left(z- z_{\rm LV}\right) Q_\circ \right]$
which propagates {\em inwards},
with the typical liquid periodicity,
$2\pi/Q_\circ$, and with a decay length $\xi_{\rm V}$ related
to the peak width $\delta Q$ in the liquid structure factor $S(Q)$,
$2\pi/\xi_{\rm V} \sim \delta Q$ \cite{noterhosl}. 
This second oscillation is not universally present: it is only
expected for a strongly relaxed and contracted surface, like that
of a metal, {\em which represents a heavy disturbance for
the liquid below} \cite{ICET}.
In this sense, the LV interface in a metal is similar to a hard wall.
By contrast, at the more disordered LJ liquid surface this oscillation 
(still present in principle)
is below 2\%\cite{LJLV} and practically irrelevant at $T_m$.
A quantitative picture obtained by simulation 
(fig.\ \ref{fig:prof}, described later) shows that a large oscillation 
is present for Au and absent in LJ.
When $\ell=z_{\rm LV}-z_{\rm SL}$ decreases,
the two density oscillations overlap, giving rise,
within linear response, to
\bq
f_{\rm L}(\ell) = \frac{1}{2} \int d^3 r d^3 r' C({\bf r}-{\bf r'})
\delta \rho_{\rm SL}({\bf r}) \delta  \rho_{\rm LV} ({\bf r'})
\label{eq:free}\eq
where the integration should be restricted to the liquid
region $ z_{\rm SL} \le z \le z_{\rm LV} $, and $C(\bf{r}-\bf{r'})$ is
the liquid density response function.
Taking as a simple approximation for large $r$ the form
$C(r)=C_\circ \exp (-r/\xi)\cos (Q_\circ r)$,
the integral can be carried out explicitly.
The (111) surface corresponds to $Q_\circ a=2\pi$,
the (110) surface to $Q_\circ a=2\pi\sqrt{3/8}$.
In both cases $f_{\rm L}(\ell)$ is oscillatory, with a wavelength
$2\pi/Q_\circ$ controlled by $C(r)$, and the lower envelope of
the minima indicates interface attraction.
However, as a consequence of tuning (commensurability), 
{\em we find that the attractive minima of (\ref{eq:free})
are about one order of magnitude
deeper in the (111) case}. Since the final surface melting behavior
is determined by $f_{\rm SR}(\ell) +f_{\rm L}(\ell)$, 
the competition between SM and NM reduces, in our picture, to a matter of:
a) how large  $f_{\rm SR}$ (generally repulsive) is;
b) how important the liquid surface oscillation  $\delta\rho_{\rm LV}$ is; and
c) how bad, or good, is the relative tuning of the two
   periodicities $2\pi/Q_\circ$ and $a$.
The binding energy between the SL and LV interfaces is given by
$\Delta\gamma \equiv \gamma_{\rm SL} + \gamma_{\rm LV} - \gamma_{\rm SV} =
  \int_0^\infty [ f'_{\rm SR}(\ell) + f'_{\rm L}(\ell) ] d\ell$.
The above result suggests that $f_{\rm L}$ contributes much more
to $\Delta\gamma$ in a well-tuned case. In a poorly packed,
poorly tuned face, the short-range repulsion
cannot easily be reversed.
In this case, fluctuations are
very effective in washing out the shallow layering minima
\cite{chernov}, whence a negative overall $\Delta\gamma$, leading to SM.
In contrast, the good tuning of a well-packed face can cause $f_{\rm L}$ to
prevail and yield the final $\Delta\gamma >0$ typical of a NM surface.

We have tested this qualitative picture
by a microscopic 1D model calculation. We extend the
Tarazona-Vicente (TV) scheme \cite{vicente}---which demonstrates
layering of a fluid near a wall---to describe a fluid
confined between a wall (the LV interface) and a density-wave solid
with adjustable periodicity.
A 1D lattice-gas fluid (occupation
$\rho_i = 0,1$, lattice spacing equal to 1),
is described by a grand canonical free energy
\bq
\Omega\{\rho_i\} = F\{\rho_i\} + \sum_i(V_i-\mu)\rho_i .
\label{eq:grand}\eq
We take
\bq
F\{\rho_i\} = k_B T \left[\rho_i\ln \rho_i +
(1-\rho_i)\ln(1-\rho_i)\right] - \alpha\rho_i\rho_{i+n}
\label{eq:effe}\eq
so as to describe a simple fluid.
The first term in (\ref{eq:effe}) is entropic, and the second is
a gradient, representing short-range order in the fluid, with
periodicity $2\pi / Q_\circ \approx n$.
The external potential $V_i$ consists of a hard wall
at $i=0$ and of a semi-infinite ``crystal'' of periodicity
$a=\lambda n$ located at some very large distance $i=L$ away.
The hard wall generates a density oscillation decaying
into the liquid from the left with periodicity $n$.
The crystalline perturbation generates a second oscillation
decaying into the liquid from the right, with periodicity
$\lambda n$, where $\lambda$ is a detuning parameter
mimicking the crystalline interplanar spacing of a general
surface orientation.
The values of all the $\rho_i$'s are determined so as to minimize
$\Omega$. As the crystal is moved closer and closer to the
wall by reducing $L$, the fluid in between is modulated with
both periodicities, and so is the excess of surface free
energy $\gamma(L)=\Omega(L) + PL$,
where $P=k_B T \ln (1-\rho) + \alpha\rho^2$
is the lattice gas pressure \cite{vicente}.
The resulting surface free energy difference
$\Delta\gamma(L)\equiv\gamma(L)-\gamma(\infty)$
in the case $\alpha/k_B T=-4$ 
(as used by TV for a strongly structured fluid)
is shown in fig.\ \ref{fig:tarazonismi},
together with density profiles.
When the two periodicities are in tune
($a=2\pi/Q_\circ$), there is a sequence of attractive
minima as a function of the thickness $L$ (upper panel).
Trapping in the deepest of these minima will correspond to NM, as discussed.
When tuning is absent (lower panel), the minima are still found, but there is
no net attraction left. Prevalence of interface repulsion
will favor SM in such a case.

The value of these concepts, if correct, should be readily
recognizable in experiments (not yet available), or at least
in a realistic MD simulation.
For this purpose,
we have conducted parallel simulations on Au(111) and Au(110),
the former being a well known NM face \cite{stock},
and the latter a SM face \cite{hoss}.
We have used the ``glue model'' many-body potential \cite{philmag},
which reproduces correctly, among other things, the bulk
melting temperature (within $\sim 1\%$), the solid-vapor \cite{philmag}
and the liquid-vapor \cite{ICET} surface energies,
the main surface reconstructions of gold \cite{philmag},
as well as the NM and SM
behavior of these two surfaces \cite{EITTC,CET,lione}.
MD simulations were done
on a variety of systems, in the form of
$N$-layer slabs ($32\leq N\leq 80$) with $M$ atoms per layer
($48\leq M\leq 280$), one free surface and lateral periodic
boundary conditions.
Special care was taken in determining the effective bulk
melting temperature, which {\em in a cell-periodic system
is both size and orientation dependent} \cite{lione}
[for example we find $T_m=1355\pm 5\,\rm K$ for a (111) bulk cell with
$N=80$, $M=56$,  and $T_m=1327\pm 3\,\rm K$ for a (110)
bulk cell with $N=32$, $M=280$.
These values are close to the actual $T_m$ of Au, $1336\,\rm K$].
Near $T_m$, we generate configurations
[which can be of equilibrium on Au(110), and near equilibrium on Au(111)],
where a certain number $n$ of surface layers are melted.
The density profiles of the two interfaces, SL and LV, are shown
in fig.\ \ref{fig:prof}.
The two oscillations, one attached to the LV interface,
the other to the crystal, are clearly visible.
As anticipated, their wavelengths are quite close for Au(111)
($2\pi/Q_\circ =2.30 \pm 0.05\,\rm\AA$, $a=2.39\,\rm\AA$),
and very different for Au(110) ($a=1.46\,\rm\AA$).

In order to obtain information on the behavior of the total free energy
as a function of $\ell$, we have developed a new method based
on non-equilibrium recrystallization runs.
Starting from a slab configuration with a large
number of melted layers,
the temperature $T$ is suddenly reduced from
an initial value $T_i > T_m$ to a final value $T < T_m$.
The SL interface moves
towards its final equilibrium position
with a velocity $d\ell/dt$ (fig.\ \ref{fig:veloc}).
As long as the liquid film is thick,
the interface motion is approximately uniform. At small thickness,
however, we find a final speedup for the (111) face but,
in contrast, a final slowdown for the (110) face.
The reverse experiment, i.e. fast melting, can also
be done, and the reverse behavior is observed: initially,  
the melting front moves out quickly on (110), slowly on (111).
By analyzing the behavior of $\ell$ as a function of time $t$,
we extract the effective SL--LV interaction free energy $\bar{f}(\ell)$,
the average now fully including interface fluctuation effects.
The driving force on the SL interface is given by
$-d{\cal F}/d\ell$, where $d{\cal F}$ is the
change in the total free energy per unit area when
the liquid thickness is changed from $\ell$ to $\ell + d\ell$:
\bq
d{\cal F} =  \rho L (1-T/T_m) d\ell
+ \bar{f}(\ell+d\ell) - \bar{f}(\ell) ,
\label{eq:freechg}\eq
where $\rho$ is the density of the liquid, and
$L$ the latent heat of melting.
Close to $T_m$, the velocity may be assumed to be
proportional to the driving force (classical body in a viscous medium):
\bq
\frac{d\ell}{dt} =
- A [\rho L (1-T/T_m) + \bar{f}'(\ell)]
\label{eq:veloc}\eq
For $\ell > 20\,\rm\AA$, $\bar{f}'(\ell)$ appears to be negligible,
and we obtain
$A_{111} = 3.0\times 10^{11}\rm \AA^4 s^{-1} meV^{-1}$ and
$A_{110} = 4.0\times 10^{11}\rm \AA^4 s^{-1} meV^{-1}$.
We checked that they do not depend on $T$
in the region of interest (within 100K from $T_m$).
By assuming $\bar{f}(\ell) = -\Delta\gamma \exp(-\ell/\xi)$,
eq. (\ref{eq:veloc}) can be easily integrated, giving
\bq
\ell (t) =  \xi \ln\left(C{\rm e}^{-t/\tau}
- \frac{\Delta\gamma}{\xi\rho L (1-T/T_m)} \right)
\label{eq:ell}\eq
where
$\tau=\xi/[A\rho L (1-T/T_m)]$
and $C$ is the integration constant, fixed by the initial liquid thickness
$\ell(0)$.  In the equilibrium limit $t\rightarrow +\infty$, and
if $\Delta\gamma<0$, eq.\ (\ref{eq:ell}) reduces to the well-known logarithmic
growth law of $\ell$ as a function of $T$ \cite{vanderveen}.
From previous equilibrium or quasi-equilibrium simulations
\cite{EITTC,CET,lione} we extract
(using also the relationship between maximum overheating temperature
and $\Delta\gamma$ \cite{ditolla})
$\Delta\gamma_{110}=-8.5\pm 3.0\,\rm meV/\AA^2$,
$\xi_{110}=2.3\pm 0.2\,\rm\AA$,
$\Delta\gamma_{111}=+7.5\pm 2.0\,\rm meV/\AA^2$,
$\xi_{111}=11\pm 1\,\rm\AA$.
Together with $\rho=0.0562\,\rm \AA^{-3}$, $L=0.112\,{\rm eV/atom}$
(known values for this potential) we calculate the theoretical
$\ell(t)$ from eq.\ (\ref{eq:ell}). 
This reproduces simulation data fairly well (fig.\ \ref{fig:veloc}), 
except where $-d{\cal F}/d\ell$ becomes small and dynamics slows down
[e.g., $\ell\lesssim 15\,\rm\AA$ for (110)].
Clearly, the overall effective interaction free energy is
{\em repulsive} in the (110) case,
and {\em attractive} in the (111) case. Its magnitude
is surprisingly large, as much as 10\% of the total surface free energy
($71\,\rm meV/\AA^2$ for liquid Au at $T_m$ \cite{ICET}).
The individual layering oscillations
are generally not observable in the recrystallization interface velocity,
washed out by the large intrinsic
interface width, and by fluctuations.
However, the large value of $\xi_{111}$ shows that 
the (111) attraction is liquid-mediated,
and must therefore come from layering, 
as also suggested by fig.\ \ref{fig:prof}.
Conversely, the small value of $\xi_{110}$ reflects
the destructive interference of the two oscillations.

In summary, we have argued theoretically, and demonstrated
both in a 1D model and realistic recrystallization simulations the
changeable role of the layering on surface melting.
The strong effect of the liquid surface density oscillations, and
their tuning-detuning effect has been demonstrated. Through X-ray detection
of the magnitude of these oscillations, now possible \cite{magnussen},
these results should shed new light on the high-temperature behavior of
solid surfaces.

We acknowledge support from
EEC under contract ERBCHRXCT930342, and from CNR under project SUPALTEMP.



\begin{figure}
\caption{
Comparison between the ($x,y$ averaged)
density profiles of the (111) and (110) surfaces
of LJ and Au, when wetted by a liquid film
of thickness $\ell\sim 30\,\rm\AA$ (or $\ell\sim 12$ LJ units),
obtained by constant energy MD simulations.
The film thickness is stabilized by energy conservation.
Layering oscillations induced in the liquid film by the solid
are present in all cases.
The metal also exhibits a strong layering
on the opposite side of the film, induced by the liquid surface.
Its wavelength is rather different from that of the solid-induced
oscillation in the case of Au(110), but very close for Au(111).
Upon releasing the constant energy constraint at $T\approx T_m$,
Au(111) is unstable and recrystallizes \protect\cite{CET}.
}
\label{fig:prof}
\end{figure}

\begin{figure}
\caption{
Variation of the surface free energies $\Delta\gamma$ (with respect to the
infinite separation limit) as a function of the interface separation $L$,
using the lattice gas models described in the text.
Upper panel: (111)-like case (in-tune perturbations).
Lower panel: (110)-like case (out-of-tune perturbations).
The lower envelope of the two curves
shows strong interface attraction for ``(111)'', and weak attraction
followed by repulsion at shorter distances for ``(110)''.
Corresponding typical plots of densities $\rho_i$ as a function of
lattice site are shown in the insets.
In the ``(110)'' case, the periodic potential induces crystal-like
density oscillations for $i>L$.
}
\label{fig:tarazonismi}
\end{figure}

\begin{figure}
\caption{
Liquid thickness $\ell$ as a function of time $t$
as obtained from non-equilibrium MD quenching simulations
from $T=1500\,\rm K$ to $T=1250\,\rm K\simeq 0.93\,T_m$
for Au(111) (filled dots) and Au(110) (open dots).
The starting point was in both cases a liquefied slab, more
than $100\,\rm\AA$ thick, sitting on top of a few (111) or
(110) rigid crystalline layers.
Curves corresponding to eq.\ (\protect\ref{eq:ell}) are shown as
a solid [(111)] and dotted [(110)] line.
Recrystallization proceeds at constant (but slightly
orientation-dependent) speed until $\ell\sim 40\,\rm\AA$.
For (111), the speed increases in the final stage (leading to a
crystalline (111) surface), indicating
attraction between the SL and the LV interface.
The opposite occurs on (110), leading to a surface melted state.
In this case the system reaches the equilibrium thickness
near $t\sim 1.2\,\rm ns$.
}
\label{fig:veloc}
\end{figure}


\begin{references}

\bibitem{argon}
  J.G. Dash, Contemp. Physics {\bf 30}, 89 (1989),
  and references therein.
\bibitem{LJ}
  S. Valkealahti and R. M. Nieminen,
  Physica Scripta {\bf 36}, 646 (1987);
  Y. J. Nikas and C. Ebner,
  J. Phys.: Condens. Matter {\bf 1}, 2709 (1989);
  X. J. Chen {\em et al.},
  Surf. Sci. {\bf 249}, 237 (1991);
  {\it ibid.}, {\bf 264}, 207 (1992).
\bibitem{vanderveen}
  J. F. van der Veen, B. Pluis, and A. W. Denier van der Gon,
  in {\em Chemistry and Physics of Solid Surfaces VII},
  R. Vanselow and R.F. Howe (Eds.)
  (Springer, Berlin, 1988), p.~455.
\bibitem{chernov}
  A. A. Chernov and L. V. Mikheev,
  Phys. Rev. Lett. {\bf 60}, 2488 (1988);
  Physica A {\bf 157}, 1042 (1989).
\bibitem{vicente}
  P. Tarazona and L. Vicente,
  Molec. Phys. {\bf 56}, 557 (1985).
\bibitem{noterhosl}
  Such a component in $\delta\rho_{\rm SL}$, hypothesized in
  ref.\ \cite{chernov}, is in practice not found
  (see fig.\ \ref{fig:prof}).
\bibitem{ICET}
  S. Iarlori, P. Carnevali, F. Ercolessi, and E. Tosatti,
  Surf. Sci. {\bf 211/212}, 55 (1989).
\bibitem{LJLV}
  D. L. Heath and J. K. Percus,
  J. Stat. Phys. {\bf 49}, 319 (1987);
  M. J. P. Nijmeijer {\em et al.},
  J. Chem. Phys. {\bf 89}, 3789 (1988).
\bibitem{EITTC}
  F. Ercolessi, S. Iarlori, O. Tomagnini, E. Tosatti, and X. J. Chen,
  Surf. Sci. {\bf 251/252}, 645 (1991).
\bibitem{stock}
  K. D. Stock and B. Grosser,
  J. Cryst. Growth {\bf 50}, 485 (1980).
\bibitem{hoss}
  A. Hoss, M. Nold, P. von Blanckenhagen, and O. Meyer,
  Phys. Rev. B {\bf 45}, 8714 (1992).
\bibitem{philmag}
  F. Ercolessi, M. Parrinello, and E. Tosatti,
  Philos. Mag. A {\bf 58}, 213 (1988).
\bibitem{CET}
  P. Carnevali, F. Ercolessi, and E. Tosatti,
  Phys. Rev. B {\bf 36}, 6701 (1987).
\bibitem{lione}
  F. Ercolessi, O. Tomagnini, S. Iarlori, and E. Tosatti,
  in {\it Nanosources and manipulation of atoms under high fields
  and temperatures: applications},
  ed. by Vu Thien Binh, N. Garcia, and K. Dransfeld,
  (Kluwer, Dordrecht, 1993), p.\ 185.
\bibitem{ditolla}
F.~D. Di Tolla, F. Ercolessi, and E. Tosatti,
Phys. Rev. Lett. {\bf 74}, 3201 (1995).
\bibitem{magnussen}
O.Magnussen {\em et al.}, Phys. Rev. Lett. {\bf 74}, 4444 (1995).

\end{references}
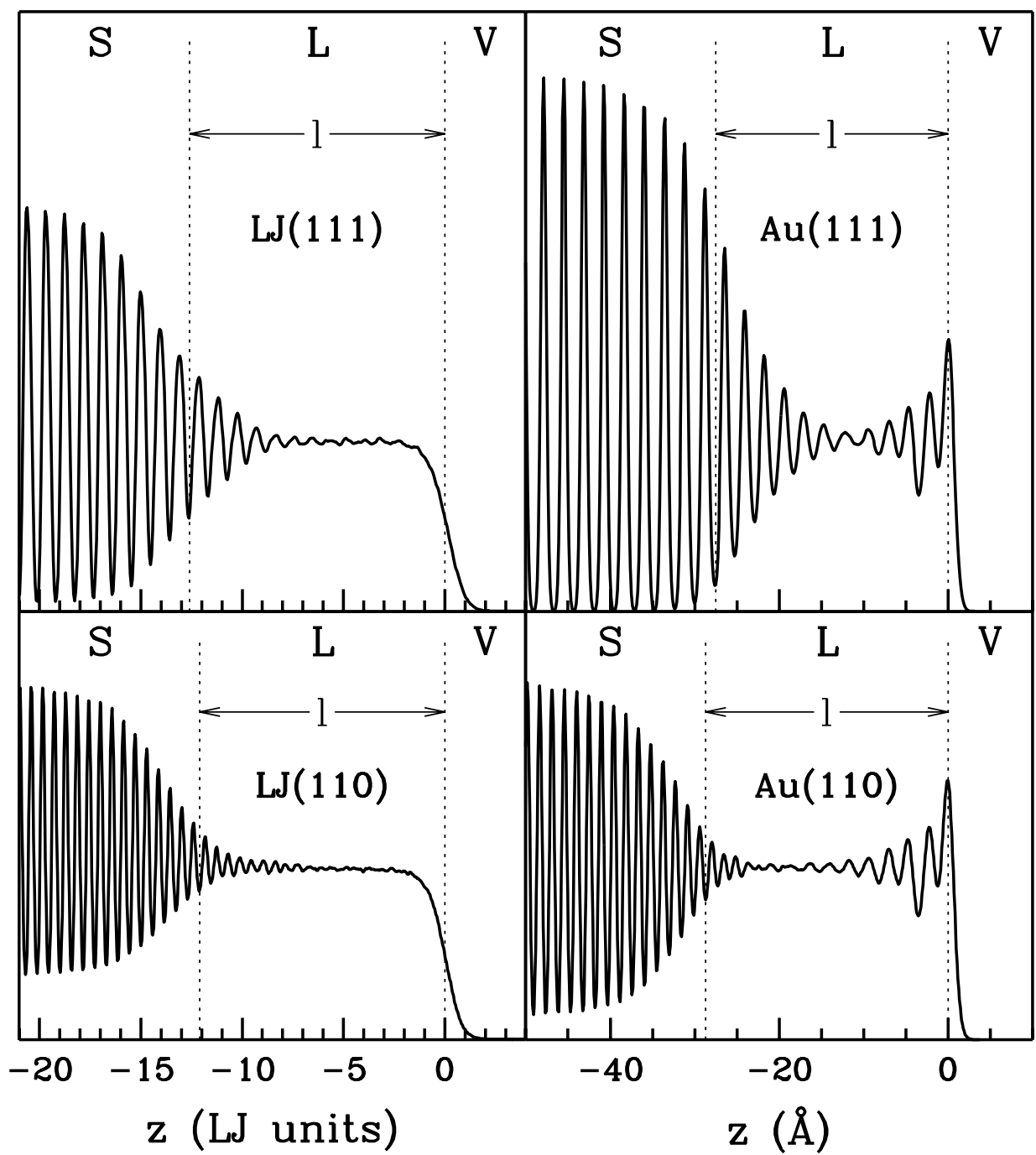
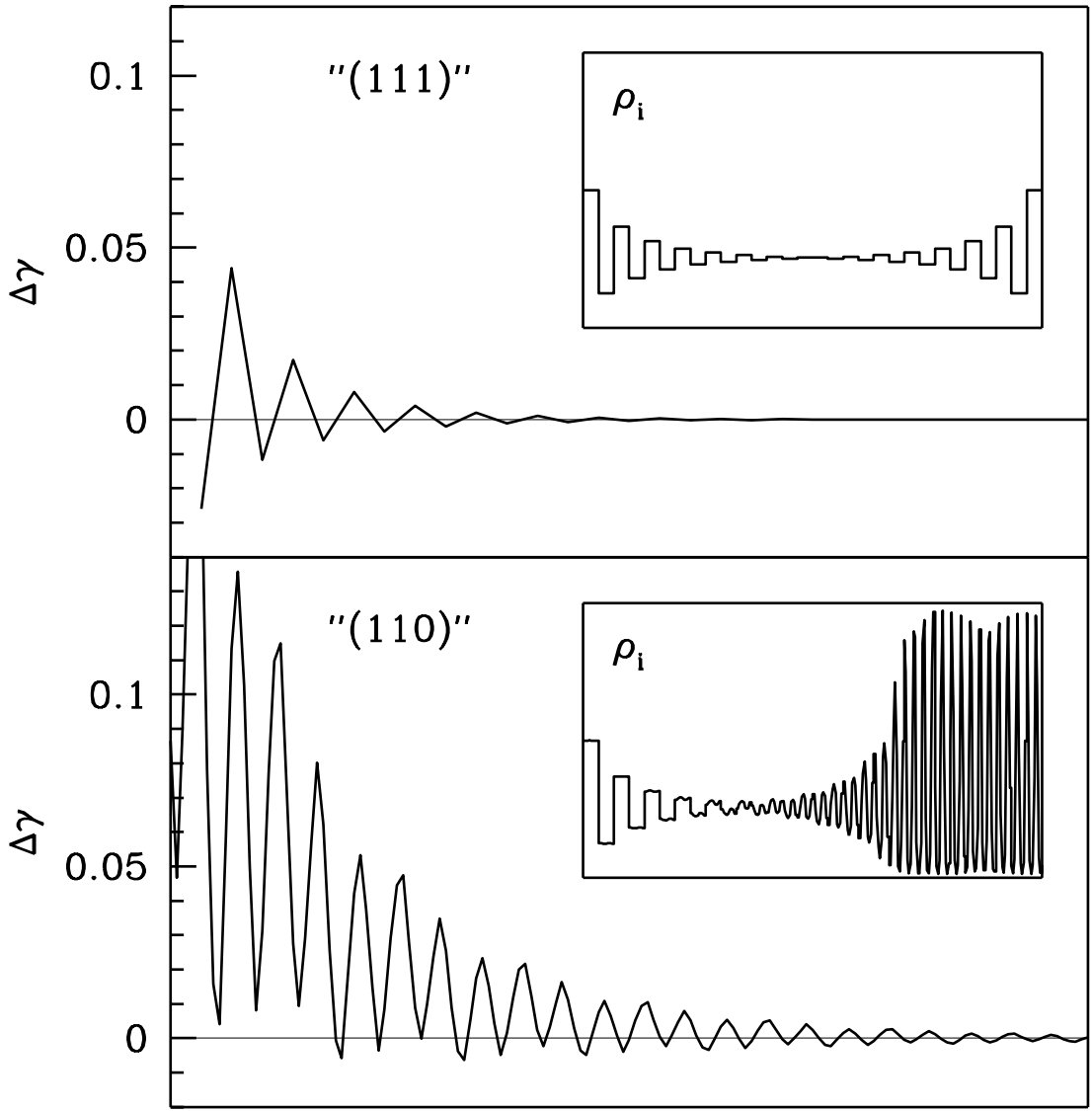
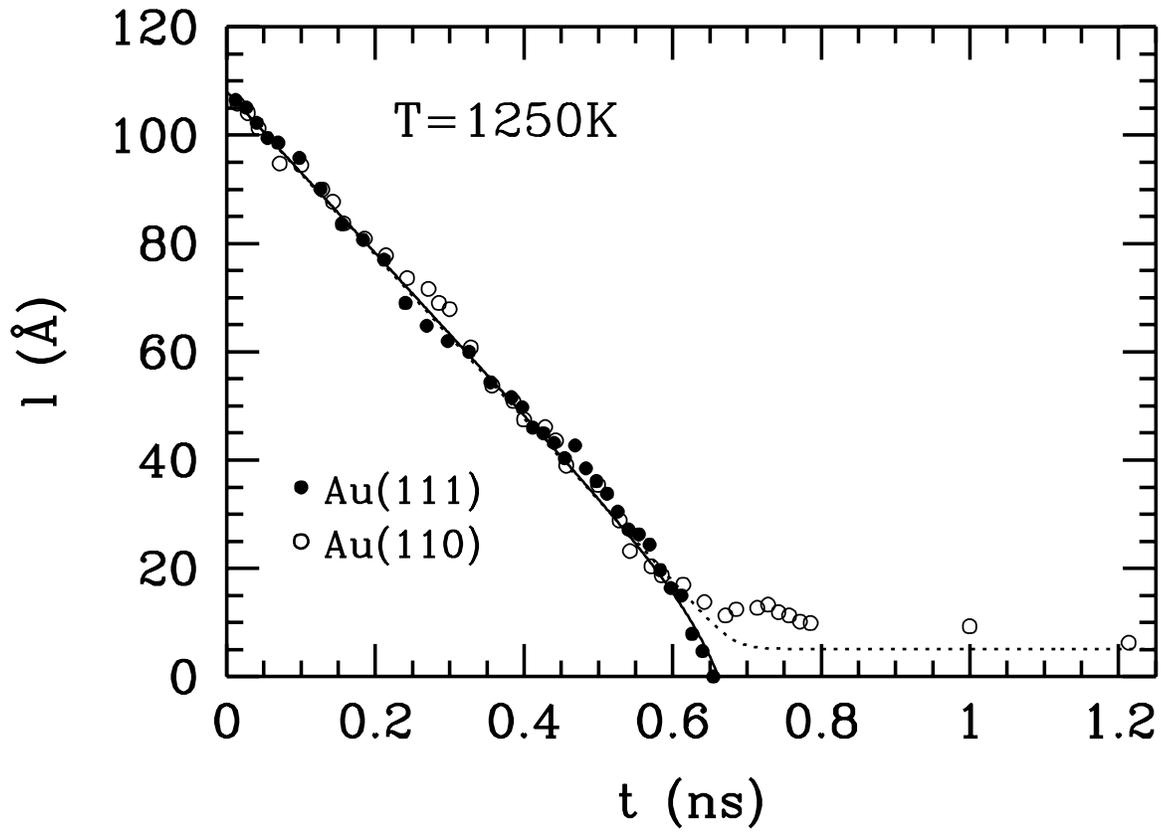
\end{document}